\documentclass[12pt,preprint]{aastex}

\shorttitle{X-Ray Absorption Halo}
\shortauthors{Bregman and Lloyd-Davies}

\begin{document}

\title{X-Ray Absorption from the Milky Way Halo and the Local Group}

\author{Joel N. Bregman and Edward J. Lloyd-Davies}
\affil{University of Michigan, Department of Astronomy, Ann Arbor, MI 48109}
\email{jbregman@umich.edu, ejdavies@umich.edu}

\begin{abstract}
Million degree gas is present at near-zero redshift and is due either to a
gaseous Galactic Halo or a more diffuse but very massive Local Group
medium.  We can discriminate between these models because the column
densities should depend on location in the sky, either relative to the
Galaxy bulge or to the M31-Milky Way axis.  To search for these
signatures, we measured the \ion{O}{7} K$\alpha$ absorption line strength toward 25
bright AGNs, plus LMC X-3, using {\it XMM-Newton RGS\/} archival data.  The data
are in conflict with a purely Local Group model, but support the Galactic Halo
model.  The strongest correlation is between the \ion{O}{7} equivalent widths
and the {\it ROSAT\/} background emission measurement in the R45 band (0.4-1 keV), 
for which \ion{O}{7} emission makes the largest single contribution.  This
suggests that much of the \ion{O}{7} emission and absorption are cospatial, from which
the radius of a uniform halo appears to lie the range 15-110 kpc. 
The present data do not constrain the type of halo gas model and
an equally good fit is obtained in a model where the gas density decreases 
as a power-law, such as r$^{-3/2}$.  
For a uniform halo with a radius of 20 kpc, the electron density would be 
9$\times$10$^{{\rm -4}}$ cm$^{-3}$, and the gas mass is 4$\times$10$^{{\rm 8}}$ M$_{\odot}$.  
The redshift of the four highest S/N \ion{O}{7} measurements is consistent with a Milky
Way origin rather than a Local Group origin.

\end{abstract}


\section{Introduction}

One of the important recent results from {\it Chandra, XMM,\/} and {\it FUSE\/} is
the detection of hot gas at nearly zero redshift.  The low redshift of the
gas implies that it is either in the Local Group or in a halo around the
Milky Way, and discriminating between these two possibilities is critical. 
If the gas is extended and fills the Local Group on the Mpc scale, its mass
exceeds the baryonic mass in all other galaxies in the Local Group.  This
would account for some of the ``missing baryons'' in the Local Universe,
of fundamental importance for cosmology.  Alternatively, if the gas is
local to the Milky Way, within a 100 kpc region, the baryonic mass is two
orders of magnitude lower and it would not dominate the local baryon
content.

The tracers for this hot gaseous medium are the lines from common ions
that occur in the 10$^{{\rm 6}}$ - 10$^{{\rm 7}}$ K range, most 
notably \ion{O}{7} and \ion{O}{8}.  These
X-ray lines have been observed in absorption \citep{nica02, rasm03} with
both {\it Chandra\/} and {\it XMM\/}, and in emission with a 
quantum-microcalorimeter spectrograph on a {\it Shuttle\/} payload, DXS
\citep{mcca02}.  In addition, the \ion{O}{6} ion, most common in 10$^{{\rm 5.5}}$ K gas,
has been studied in over 100 sightlines with {\it FUSE\/}
\citep{semb03,wakk03}.

The location and mass of the hot gas depends strongly on the assumed
relationships between the ions.  If \ion{O}{6}, \ion{O}{7}, and \ion{O}{8} are cospatial,
\citet{nica02} cannot find a pure collisional model at a single temperature.
However, he can fit his data with a low density medium where photoionization is important,
with a temperature of 2.5$\times$10$^{{\rm 5}}$ K, a gas density of 
n$_{{\rm e}}$ $\approx$ 6$\times$10$^{{\rm -}{6}}$ cm$^{{\rm -}{3}}$, and
a pathlength of 3 Mpc, corresponding to a gas mass 
of 1.7$\times$10$^{{\rm 13}}$ M$_{{\rm \odot}{}}$.  This gas mass is 
considerably greater than the mass of all stars and cold gas within this
volume, which is about 2$\times$10$^{11}$ M$_{\odot}$ and 
1.2$\times$10$^{10}$ M$_{\odot}$, respectively \citep{braun91,cox00,vand00,cari06}.  
This ``Local Group Model'' would help explain the missing
baryons in the local universe.

Alternatively, \citet{rasm03} assume that only the X-ray lines are
cospatial, primarily \ion{O}{7} and \ion{O}{8}, leaving out \ion{O}{6}.  They fix the
temperature from the \ion{O}{8}/\ion{O}{7} ratio, and, including the emission
observations, find that a collisional ionization model is possible with a
length scale of 0.15-1 Mpc.  Their result implies a Local Group medium
with masses of 10$^{{\rm 13}}$ M$_{{\rm \odot}{}}$ for the 1 Mpc 
length scale, and 10$^{{\rm 11}}$ M$_{{\rm \odot}{}}$ for the
0.15 Mpc length scale, still more gas than the collective ISM of Local
Group galaxies.  Recently, \citet{fang05b} pointed out the sensitivity of
this result to temperature and for a somewhat lower permitted
temperature, smaller length scales are allowed, typical of a Galactic halo
(50 kpc), and leading to considerably reduced gas masses.

In yet another approach, one might assume that the \ion{O}{7} emission and
absorption are from the same gas, with some of the \ion{O}{8} and \ion{O}{6}
originating in other material.  The X-ray emission line data was obtained
with the inaugural flight of a quantum calorimeter rocket flight by
\citet{mcca02} where they detected individual spectral lines responsible
for much of the soft X-ray background, leading to an emission measure
and metallicity.  With an emission measure (n$^{{\rm 2}}${\it L\/}) and an absorption column
(n{\it L\/}), the size of the region {\it L\/} can be obtained and 
this leads to {\it L\/} $\approx$ 50 kpc
\citep{sand02}, although this is a lower limit if one allows for filling
factors below unity.  For filling factors near unity, this solution describes
a halo around the Galaxy where the gas density is 
n$_{{\rm e}}$ $\sim$ 10$^{{\rm -}{4}}$ cm$^{{\rm -}{3}}$ and the
hot gas mass is $\sim$10$^{{\rm 9}}$-10$^{{\rm 10}}$ M$_{{\rm \odot}{}}$, 
2-3 orders of magnitude lower than the
solutions of Nicastro or Rasmussen.  In this ``Galactic Halo'' solution, the
\ion{O}{6} is not cospatial, as argued by \citet{sav05a} and \citet{will05}, but might arise
nearby as the \ion{O}{7} gas cools and recombines or if there is a conductive
interface in a HVC.

The ``Local Group'' and ``Galactic Halo'' models represent limiting cases. 
The hot gas cannot be too small (e.g., in the disk) or it would
overproduce the soft X-ray background.  Also, the absorbing region
cannot be larger than the virial radius of the Local Group or the gas
would not have a potential well to fall into and become heated.  To
distinguish between these possibilities, one can search for an angular
signature to the absorption line strengths.  In particular, for a Local
Group model, the absorption should be more pronounced for sight lines
closer to the Milky Way - M31 axis.  For a Galactic Halo model, the
absorption should be more pronounced for sight lines across the Galaxy,
compared to those away from the galaxy.  Searching for such signatures
requires a statistically significant number of sightlines with measured 
X-ray absorption lines, and for this purpose, the \ion{O}{7} resonance absorption
line is the easiest to detect.  There are relatively few {\it Chandra\/}
observations of this line toward bright sources, in part because the
gratings are not frequently used (but they have still proven quite useful;
\citealt{mcker04,fang05b,yao05,wang05}).  However, for {\it XMM\/} observations, the combination
of the spectrograph ({\it RGS\/}) always operating plus the larger collecting
area has led to a significant number of viable targets for \ion{O}{7} absorption
measurements.  Here we present the sample of the brightest AGNs that
have been observed by {\it XMM\/} and were available through the archive as of
April 2006.  We show that this sample leads to a compelling case for the
Galactic Halo model of X-ray absorbing gas.

\section{Object Selection and Data Analysis}

We wish to sample a path length as large as the Local Group, which
implies the selection of extragalactic sources, with AGNs being the most
suitable.  The only AGNs that can be used for X-ray absorption line
studies are bright and have been observed for a sufficiently long time.  We
constructed a list of the brightest AGNs from the archival data obtained
by {\it ROSAT\/} and {\it ASCA\/}. This is a valuable starting point, but AGNs are
variable, so some sources that were fainter in the past can be brighter
now.  Therefore, using the archive, we inspected the {\it XMM-Newton RGS\/}
data for dozens of targets and chose the most suitable objects.  The
quantitative goal was to find sources for which the resulting uncertainty
in the \ion{O}{7} equivalent width is about 10 \AA\ or less, since the typical
absorption line strength was seen to be about 20 \AA\ in previous studies. 
This led to a sample of 25 AGNs, plus we also included a source in the
LMC, LMC X-3 (Table 1).

Many of these sources have been observed multiple times over the
lifetime of XMM and so the data can be combined to improve the
signal-to-noise. Unfortunately CCD 4 in the RGS2 failed shortly after
launch and since it covers the wavelength range 20.1$-$23.9 \AA\  only the
RGS1 data is useful.  The data was reduced in the usual manner, using
the XMM SAS software (version 6.5), and was cleaned using an interative three sigma
clipping of a lightcurve from a background region on CCD 9, to remove
periods of high background that would degrade the signal-to-noise. 
First order source and background spectra were produced from the 
cleaned events using the SAS task RGSSPECTRUM and response matrices 
were generated using RGSRMFGEN.
The RGS1 first order spectra from the observations of each source were
background subtracted, and after each spectrum was corrected for the 
effective area, they were combined to produce a single spectrum for each source.
We note that the instrumental redistribution of photons in channel space is not corrected
for. This is considered below.

To fit absorption lines, we excluded regions of known detector features
and regions of intrinsic AGN absorption and emission lines.  For the \ion{O}{7}
resonance line at 21.603 \AA , there is a detector feature at about 21.82 \AA\ 
when the data are reduced with the standard SAS, so the 21.77-21.84 \AA\ region
was excluded in the fits (blank regions in our figures).  Many of these bright
objects are nearby Seyfert galaxies, which have their own \ion{O}{7} lines,
some of which occur in absorption and some in emission.  In most cases,
the lines are sufficiently redshifted that they do not contaminate the local
\ion{O}{7} absorption, but they restrict the region redward of 21.6 \AA\ that
can be used to define the baseline.  The redshifted lines are marked on the
figures of the spectra when relevant: MCG-6-30-15, Mkn766, NGC 3783,
NGC 4051, NGC 5548, NGC 3516, NGC 4593, and IC4329a (Figures 1-6).  In the few cases where
there are no useful continuum regions redward of 21.6 \AA , we used a flat 
continuum to the blue side of the line.  There were objects for 
which the intrinsic absorption or emission line is badly blended with the 
Galactic \ion{O}{7} absorption line, such as NGC 4151.  Although it is suitably bright, 
it has a broad \ion{O}{7} resonance emission line that overlaps with the z = 0 \ion{O}{7}
feature.  The local absorption line, although present, lies in the sharply
rising wing of the NGC 4151 emission line, so the equivalent width of the
absorption feature cannot be measured reliably. NGC 4151, and other similar
objects were not included in the sample.

When there are no intrinsic AGN features near 21.6 \AA , we fit a linear
continuum in the range 21.2 - 21.96 \AA , the upper limit being set by the
need to avoid Galactic \ion{O}{6} K$\alpha$ absorption at 22.05 \AA\ \citep{prad03}.
In addition to the continuum, we fit a Gaussian absorption line for the 
Galactic \ion{O}{7} absorption feature.  The least-squares minimization has
been performed by both the Simplex method and using a Levenberg-Marquardt
minimization and we arrive at the same result.  In practice, there are five parameters
in this model:  the continuum level; the slope of the continuum; the central
wavelength of the absorption line; the Gaussian $\sigma$ of the absorption line;
and the area of the line (the line depth), which can be negative or positive
(absorption and emission are permitted).  Certain constraints can be applied
to some of these parameters.  For example, the Gaussian fit cannot be narrower than the 
instrumental width ($\sigma$ = 0.03 \AA\, or FWHM = 0.06 \AA ) and it probably will not be wider than 
a velocity corresponding to the escape velocity of the Local
Group, which we estimate to be about 700 km s$^{-1}$, or $\sigma$ = 0.05 \AA .
Similarly, the location of the line center should not be outside of sensible
Local Group velocities, which sets a displacment of about 500 km s$^{-1}$,
or 0.036 \AA .  In carrying out our fits, we calculate an uncertainty for each
parameter.  Only in a few cases are the uncertainties for the line 
center and line width smaller than the allowed range, 
and for the five highest S/N sources, Mkn 421, PKS 2155-304, 3C273, MCG-3-30-15, and LMC X-3, 
the median values of the central wavelength and line width are 21.605 \AA\,
and $\sigma$ = 0.081 \AA, with small ranges.  Consequently, for fainter sources
where the central wavelength and line width cannot be accurately determined, we
fix the central wavelength at 21.603 \AA\ and the line width at $\sigma$ = 0.08 \AA.
When we relax these constraints, the values for the equivalent widths do not vary
greatly, although the uncertainty will naturally be larger. We note that the uncertainty
determined from photon counting statistics for the bins of the combined spectra
appear to overestimate the true 
rms since the least square fits have a $\chi$$^{{\rm 2}}$ that is too small (one
can verify this by inspection in that a fit goes through nearly every error bar).
To achieve a typical $\chi$$^{{\rm 2}}$ for an acceptable fit, we estimate that the
uncertainty per point would need to be reduced by 25-80\% , depending on
the object.  This is due to a misalignment between the raw channels and the flux
binning, introduced in the creation of the combined spectra, resulting in a lack of 
independence between adjacent flux bins.  To verify that neither the non-independence of
the flux bins nor the lack of treatment of the instrumental redistribution are
significantly affecting our results, we have simultaineously fitted 
the individual spectra in XSPEC for some test cases.

Upon examining the spectra, there were no absorption features that deviated from
expectations, with the exception of NGC 7469, where there is a possible
absorption feature at -900 km s$^{{\rm -1}}$ (98\% confidence), a velocity shift is
much greater than the instrumental uncertainty.  However, with 26
objects, there is a near-unity chance of finding one such feature over a
range of $\pm$ 1000 km s$^{{\rm -1}}$, so we do not consider this a real detection.

The resulting line strengths and statistical uncertainties are given in Table 2 and the
spectra are given in Figures 1-6.  The objects are ordered by decending values
of their continuum S/N, which is a figure of merit for the quality to which
uncertainties on the equivalent widths can be obtained.  Of the 26 objects, 10 have detections
above a 3$\sigma$ threshold, six are at the 2-3$\sigma$ level and the remainder are below 2$\sigma$. 
Even non-detections are important, provided that the uncertainties are
adequately small.  For this sample, 18 objects have uncertainties to the
equivalent widths of 10 \AA\ or less, and this forms the most useful subset of
our data.  There happens to be a decrease in quality of the spectra after object 18,
but we also have used the remaining objects in some tests (although as those tests
are weighted by the errors, these last eight objects do not have a large influence
on the results).

A few sources were observed with both {\it XMM\/} and {\it Chandra\/}, and in nearly
every case, the {\it XMM\/} data is of higher S/N.  The exception is the source
LMC X-3, where the {\it XMM\/} and {\it Chandra\/} data \citep{wang05} are of
comparable S/N and their equivalent widths are nearly identical.
The line strengths measured by others for the {\it XMM\/} data give very
similar results (e.g., \citealt{rass06}), the differences being consistent 
with different fitting methods.  Other approaches to fitting the same data 
(e.g., within XSPEC or IRAF) yielded nearly identical results.

There are other, weaker absorption lines present, although there are not
enough of them to form a useful set for angular studies.  The most
relevant is the \ion{O}{8} K$\alpha$ line, which can be 
used with the \ion{O}{7} K$\alpha$ line to
constrain the temperature.  The only useful objects are Mkn 421, PKS
2155-304, and 3C273, reported on by \citet{rasm03}, so we use their
results for the \ion{O}{8} to \ion{O}{7} line ratios.

\section{Analysis of the Data}

The primary goal is to determine whether the absorption lines have an
angular signature that correlates with either the Local Group structure, or
various Milky Way properties.  Given the limits of the data, it is
impractical to fit multi-parameter models to the data, so we choose simple
representative models, although we have explored a wide range of
models.

\subsection{Local Group Models}

We begin with the model for the Local Group, and as there are no widely
accepted models for the distribution of diffuse hot gas, here we assume
that the diffuse baryons are aligned along the axis defined by the Milky
Way and M31, which are the two dominant mass components of the
Local Group.  If the center of mass of the diffuse baryons lies halfway between
M31 and the Milky Way, then the greatest column density should occur in
a sightline toward M31 and the smallest column would occur in the anti-M31 
direction.  Consequently, a general test is whether the column
density decreases with increasing angle from M31.  This
prediction is not confirmed by the data (Figure 7), where the equivalent
widths weakly increase with increasing angle from M31, rather than decrease.

With a more detailed model, one can predict the expected distribution of
the column densities for comparison to the data.  The specific model that
we adopt is where M31 and the Milky Way lie at the two foci of an
ellipsoid in which the gas density is uniform.  The model is completely
defined by choosing a semi-major axis, and we show the results for a
model where the semi-major axis is twice the separation between M31
and the Milky Way (semi-major axis of 1.54 Mpc).  The normalization of
the column density of the model is a remaining parameter, which we
choose to minimize the $\chi$$^{{\rm 2}}$.  Inspection of the model fit (Figure 8)
indicates it is a poor fit, and as the reduced $\chi$$^{{\rm 2}}$ = 5.5, 
it is an unacceptable fit.  For semi-major axis from 0.8-2 Mpc, all models produce
unacceptable fits with no positive correlation coefficients, showing that it
is unlikely that the \ion{O}{7}-absorbing gas is distributed through the Local
Group.

We have also compared the location of the absorbers with the individual
Local Group galaxies to determine if they have a specific signature.  For
example, if individual galaxies were to have their own extended gaseous
halos, there would be stronger absorption lines for sight lines near those
galaxies (Figure 9).  If galaxy gaseous halos scale with galaxy size, then the largest
external galaxy halo should be around M31, where the closest sightline is
Akn 564, 29$\degr$ away, or a closest approach of 380 kpc.  The
equivalent width in this sight line (12.3 $\pm$ 4.6 m\AA ) appears to be less and
is certainly not substantially greater than the median equivalent width. 
This implies that if M31 has a significant gaseous halo, it is less than 380
kpc in radius.

The next largest galaxy with a nearby sightline is M33, where 3C 59
is 9$\degr$ away, or a 130 kpc impact parameter.  The \ion{O}{7} equivalent
width, 60.9 $\pm$ 19.2 m\AA , has a rather large uncertainty, so it is difficult
determining whether there a hot halo exists around M33.  Two dwarf
galaxies have nearby sightlines:  DDO 210, where Mrk 509 is 3$\degr$
away (42 kpc); and Leo I \& II, where Ton 1388 is 3.5$\degr$ away (14
kpc).  The equivalent width for Ton 1388 is above average but fairly uncertain, 
34.5 $\pm$ 15.7 m\AA , and is within 1$\sigma$ of the median value for the sample.
For Mrk 509, the detection is quite good (25.9 $\pm$ 7.3 m\AA ) and it also 
is within 1$\sigma$ of the median value.  So for Leo I \& II, there
appears to not be a gaseous halo either, although this dwarf pair does
not have HI nor does it have star formation \citep{mateo98}.  The
observations, which are only of high quality for M31 and Leo I \& II,
argue against substantial halos with a size an order of magnitude larger
than the optical galaxy.  The data do not rule out smaller galactic halo
that are comparable or a few times larger than the host galaxy.

There is one useful source within the LMC, LMC X-3 and it lies in the
disk, 6.9 $\deg$ (6.7 kpc) from the center of the LMC.  If a LMC halo
surrounds the galaxy, this sightline would sample half of the LMC halo 
plus the intervening gas between it and the Sun.  The source LMC X-3 has 
an equivalent width that is very close to the median value (21.0 $\pm$ 
5.0 m\AA ) and the line width is no wider than average (0.063 $\pm$ 0.012 \AA ), 
so there is no indication of additional absorption toward this source 
caused by a LMC halo.  Finally, the line center (21.592 $\pm$ 0.016 \AA ) is
consistent with the velocity of the Milky Way, but 2$\sigma$ lower in redshift 
than the LMC, which would have produced a line at 21.623 \AA .  This is
similar to the conclusions of \citet{wang05}, who used {\it Chandra\/} data 
to observe LMC X-3.

\subsection{Galactic Halo Models}

Another model that we consider is for a hot Galactic Halo, where the
simplest assumption is that a hot gas halo is centered on the Galaxy
center.  In this picture, the column density of hot gas should be a
decreasing function of angle away from the Galactic center, and as shown
in Figure 10, there appears to be the expected inverse 
correlation.  An unweighted linear
y-on-x fit shows that there is an inverse correlation with angle at the 95\%
confidence level, supporting the Galaxy Halo model.  To be more
quantitative, yet to keep the model simple, we calculate relative column
densities for a uniform sphere of radius R$_{{\rm H}}$, which is the only free
parameter (the equivalent width is constrained to be zero for zero path
length); we use 8.5 kpc as the distance of the Sun from the center of the
Galaxy.  A fit of similar significance is found (Figure 11), where a  best-fit 
$\chi$$^{{\rm 2}}$ model leads to R$_{{\rm H}}$ = 20.6 kpc, and the 95\%
confidence bounds are 15-110 kpc.  For the best-fit model, the $\chi$$^{{\rm 2}}$ is
unacceptably large, which can occur if the statistical uncertainty in the
equivalent widths are smaller than the variation due to structure in the
absorbing medium.  To obtain an acceptable value of $\chi$$^{{\rm 2}}$, we added 
an intrinsic scatter in the equivalent widths attributable 
to the structure of the absorbing gas of 4 m\AA , or 23\% of the
median value of the equivalent width.

The above model represents a spherical halo where the density {\it n\/} $\propto$
{\it r$^{{\rm 0}}$\/}, but we can also consider models where the density is a decreasing
function of radius, such as {\it n\/} $\propto$ $(1+(r/r_{c})^{2})^{-3 \beta /2}$, where $r_{c}$ is the core
radius.  The value of $r_{c}$ in most other galaxies with hot gas is less than 1
kpc, which is significantly smaller than the radius of closest approach for
all objects in the sample.  Therefore, the value of $r_{c}$ is not relevant
for our sample and we can write the density function as a power-law, {\it n\/}
$\propto$ $r^{-3 \beta\ }$, and we choose to examine {\it n\/} $\propto$ $r^{-3/2}$, as this is similar to
that found for the hot gas in ellipticals ($\beta$=1/2 in the ``beta'' models used
for ellipticals, e.g., \citealt{osull03}).  While this leads to a gas mass that
increases with radius as r$^{3/2}$ and an emission measure that increases 
logarithmically with radius, the column density is convergent at large radius.  
There is only a modest correlation between the theoretical and observed
equivalent widths, significant at the 92\% confidence level (Figure 12). This model is closely
associated with the Galaxy as well.  
From the available data, we cannot distinguish between the uniform model
and this ``beta'' model.  More complicated models can
certainly be constructed, but given the quality of the data, we do not
believe it is warranted at this time.

If this absorption is associated with a Galactic Halo, it might be correlated
with the soft X-ray background of the Galaxy.  The 1/4 keV X-ray
background is dominated by the gas within a few hundred pc of the Sun,
plus a flattened hot halo of height 2 kpc \citep{snow97}.  
This is not the X-ray emission that we expect the absorption features to be related to and
there is no correlation between the 1/4 keV surface brightness and the
\ion{O}{7} equivalent widths.  At higher energies, such as the 3/4 keV band,
the emission is a mixture of emission from hot Galactic gas plus emission
from unresolved AGNs \citep{gilli01}.  Also, the absorption of the X-rays
by neutral Galactic gas is less of a problem at these energies.  It is
probably the best energy at which to detect diffuse Galactic emission and
it has been known that there is enhanced soft X-ray emission in the
general direction of the bulge \citep{snow00}.  A correlation
exists between R45 and the \ion{O}{7} equivalent width
at the 99\% confidence level (weighted or unweighted fit), without
requiring that the fit go through the origin (Figure 13).
This relationship is better than the correlation found
with angle from the Galactic Center or with either of the above Galactic
Halo models.  For this fit, the $\chi$$^{{\rm 2}}$ = 45 (16 dof), which is unacceptable at
the 99\% confidence level, so we added a additional
rms in the equivalent width, as described above.  For an additional rms of
3.5 m\AA\ (19\% of the median equivalent width), the  $\chi$$^{{\rm 2}}$ is reduced to 
an acceptable 28.5
and the probability of a weighted fit is reduced to the 98\% confidence level.
The high initial $\chi$$^{{\rm 2}}$ is due to one low point and if this 
outlier were removed, the quality of the fit would improve.
By using the entire sample and creating error-weighted bins (Figure 14), we 
obtain a similar level of significance, although a somewhat lower
additional rms is required in this case (3.0 m\AA ) to obtain 
acceptable values of $\chi$$^{{\rm 2}}$.

The R45 band emission contains not only large-scale Galactic emission,
but individual bright features due to multiple supernovae, such as the
North Polar Spur.  We examined the location of the objects to determine
if they lie in bright regions of these features, but there are hardly any
coincidences.  The quasar 3C273 is located near a fainter part of the
North Polar Spur, but it's \ion{O}{7} equivalent width is not remarkable (24.6
$\pm$ 3.3 m\AA ).  The strongest component of the R45 emission is the \ion{O}{7}
emission triplet, so the relationship between the R45 emission and the
absorption equivalent widths suggests a correlation between the \ion{O}{7}
emission and absorption.  This implies that the \ion{O}{7} emission and
absorption is largely cospatial.

Another piece of evidence for the Galactic Halo picture is the mean 
redshift of our four highest S/N AGNs.  There is a random uncertainty
in the wavelength calibration of 0.008 \AA (XMM-Newton Users's Handbook), or 111 km s$^{-1}$, so it
is helpful to average together several objects.  Using these four AGNs,
the unweighted mean central wavelength is 21.608 \AA (+70 km s$^{-1}$), and the inferred 
standard deviation of a single point is 0.006 \AA , but to be conservative, we assume that
it is 0.010 \AA , as this is the median value for the uncertainty in the
central wavelength from individual fits.  Then, the uncertainty in the mean
central wavelength is 0.005 \AA (55 km s$^{-1}$), which places this mean about 1$\sigma$ of
the Milky Way (z = 0, or 21.603 \AA ).  The Local Group is probably at
-250 km s$^{-1}$, a wavelength of 21.585 \AA , 4.6$\sigma$ different from the 
measured mean value.  This suggests that the absorption is most likely
associated with the Milky Way and not the Local Group.

The conversion of the \ion{O}{7} equivalent width to a hydrogen density,
which will yield a length scale as well, requires knowledge of the oxygen
column density, the oxygen abundance, and the fraction of oxygen in the
form of \ion{O}{7}.  When the optical depth of the absorption line is low, there is
a simple linear relationship between the equivalent width and the column
density, N(\ion{O}{7}) = 3.48$\times$10$^{{\rm 14}}$ EW, where the equivalent width (EW)
has units of m\AA\ and the column density, N(\ion{O}{7}) has units of cm$^{-2}$.  
If the lines are resolved or if the gaseous medium is turbulent at the sound
speed, then the optical depth is small, but if the lines have a Doppler width
for oxygen (no turbulent or significant ordered motion), the optical depth 
at the line center can be 0.1-0.4.  As the optical depth rises, the higher order
lines, such as \ion{O}{7} K$\beta$, become of equal strength to the K$\alpha$ line.
We examined the strength of the K$\beta$ for the highest S/N spectrum, of Mkn 421,
and find a value of 3.6 $\pm$ 1.1 m\AA , although an instrumental feature 
contaminates the red side of the line.  The K$\beta$ to K$\alpha$ ratio is 
0.30 $\pm$ 0.11, which is within 1$\sigma$ of the ratio of the f values, 0.21.
This is consistent either with a small or modest optical depth.  Here we will
calculate column densities in the limit of low optical depth.

The ionization fraction can be determined by forming the ratio of \ion{O}{7} with
an adjacent ionization state, such as \ion{O}{6} or \ion{O}{8}, provided that the
different ionization states are cospatial.  The relationship between \ion{O}{7}
and \ion{O}{6} is poor and both \citet{sav05a} and \citet{will05} argue that the two are probably not
closely related.  It is tempting to relate \ion{O}{8} to \ion{O}{7}, and a correlation
between these two quantities would lend support to this possibility. 
Unfortunately, there are only a few objects for which we can determine
useful measurements of \ion{O}{8} absorption, the errors are significant, and
there is no statistically meaningful correlation between the two quantities \citep{rasm03}. 
It is possible to have a multi-temperature medium where the hotter
component leads to \ion{O}{8} while a slightly cooler component dominates
the \ion{O}{7}.  As the above study showed that \ion{O}{7} emission and absorption
are correlated, we will use these two measurements to obtain densities
and length scales, later considering whether it is consistent with the \ion{O}{8}
to \ion{O}{7} column density ratio.

For a Galactic Halo model, we expect the metallicity to be similar to the
value in the ISM and the Sun.  For oxygen, the ``Solar'' abundance has
changed considerably in the past two decades and is still not a settled
issue.  The most commonly cited Solar abundances come from \citet{ande89}, 
where log{\it N$_{{\rm O}}$\/} = 8.93, but \citet{holw01} recommends a value of
8.736, a 35\% reduction from the older value.  \citet{aspl05} advocates a
value at least 20\% lower, but here we adopt the value 8.74 for the oxygen
abundance.  For an ion fraction {\it f\/} = 0.5, this leads to an electron column of
{\it N$_{{\rm e}}$ = n$_{{\rm e}}$L = 5.8$\times$10$^{{\rm 19}}$ 
(N$_{OVII}$/10$^{{\rm 16}}$ cm$^{{\rm -2}}$) (f/0.5)$^{{\rm -1}}$\/} 
cm$^{-2}$, where {\it n$_{{\rm e}}$\/} is the
electron density and {\it L\/} is the path length for a constant density slab.  The
emission measure is available from only one line of sight, toward {\it l,b\/} = 90$\degr$,
60$\degr$ and with a 1 sr field of view \citep{mcca02}.  Toward this general
location, R45 is typically 200 cnt s$^{-1}$ arcmin$^{-2}$, above the mean of the lines of sight we
have studied.  For sight lines at similar values of R45, the typical
equivalent width is 25 m\AA , or a column density of {\it N$_{OVII}$\/} 
= 8.7$\times$ 10$^{{\rm 15}}$ cm$^{-2}$ and 
{\it N$_{{\rm e}}$= 5.0$\times$10$^{{\rm 19}}$\/} cm$^{-2}$.  The emission measure at Solar
metallicity is EM = n$_{{\rm e}}$$^2$L = 0.009 cm$^{-6}$ pc \citep{mcca02}, so one can use
this, adjusted to our oxygen abundance, and the column density to 
solve for the length scale {\it L\/}, and we find {\it L\/} =
19 kpc and {\it n$_{{\rm e}}$\/} = 9$\times$10$^{{\rm -4}}$ cm$^{-3}$.  
This length scale supports the Galactic Halo model for the hot
gas distribution.  Aside from the temperature dependence, 
{\it L\/} $\propto$ {\it f\/}$^{{\rm -2}}$ {\it (O/H)\/}$^{{\rm -1}}$, 
so for {\it f\/} = 1, the halo could be as small as 5 kpc.
A value smaller than 5 kpc is ruled out because it would have the consequence
of raising the emission measure above the observed value.

The \ion{O}{8} to \ion{O}{7} column density ratio is consistent with a gas whose
temperature is logT = 6.2-6.3.  This column density ratio changes rapidly
in this temperature range, so even a modest amount of temperature variation along
the line of sight (e.g., 20-30\%) can make a difference in this ratio by a
factor of several, as \citet{fang05b} points out.  The analysis of the emission
in the R45 band indicates a temperature of logT = 6.0-6.1 \citep{snow00,mcca02}, 
so the \ion{O}{7} absorption may also be occurring in a
temperature a bit too low to produce the \ion{O}{8} absorption as well. 
Temperature variation at this level is seen in some early-type galaxies
(e.g., NGC 4636; \citealt{jones02}), and if a positive temperature gradient
is present in the Milky Way halo, there could be radial regions where
\ion{O}{7} is dominant and more distant regions where \ion{O}{8} is dominant.  A
radial decomposition of the \ion{O}{7} to \ion{O}{8} ionization fraction ratio is
possible in principle, but the present data are inadequate for such work.

\section{Summary and Final Comments}

We sought to determine whether the X-ray absorption, primarily in the
\ion{O}{7} resonance line, were due to a Local Group medium or a hot
Galactic Halo.  It is possible to address this question by searching for the
spatial signature of these two models, which are distinct.  The Local
group is expected to be elongated along the axis defined by the Milky
Way and M31.  For our model, where the two galaxies are placed at the
foci of an ellipsoid of rotation about the MW-M31 axis, the greatest
column would result toward M31 and the smallest in the opposite
direction.  For the Milky Way Halo model where the halo is comparable
to the size of the Galaxy, the greatest columns would occur when looking
across the Galaxy, in the general direction of the bulge.  By good fortune,
these two models are nearly orthogonal in that the minimum direction for
the Local Group model, the anti-M31 direction, is at {\it l, b\/} = 
301$\degr$, 21.6$\degr$, where the Galactic Halo model would predict 
an excess compared to the mean sightline.

The comparison between the absorption line data and the models 
shows that the Galactic Halo model is preferred.
The data do not permit us to accurately determine the properties of the halo
so we adopted the simplest possible model, a uniform spherical halo.
In the context of this model, the characteristic radial size lies in
the range 15-100 kpc.  The redshift of the 
lines is consistent with a Galactic origin rather than a Local Group origin.  
The absorption line data correlates best with the R45 (0.4-1 keV) emission,
for which the biggest contribution is the \ion{O}{7} emission triplet.  This
suggests that the \ion{O}{7} emission and absorption are cospatial to some
degree, and for a cospatial model, we derive an independent measure of
the size of the halo, which is about 20 kpc.  Constraints on the
temperature may be provided by using the \ion{O}{8} absorption as well,
which leads to a temperature of logT = 6.1, as discussed previously 
\citep{mcca02, fang05b}.  These results are in good agreement with
the work of \citet{fang05b}, who also argue for a Galactic origin based
on {\it Chandra\/} obervations and on the absence of X-ray absorption
by other groups.

As an example of a possible halo, when we consider a 
uniform Galaxy halo of radius 20 kpc, the mass for 
the hot gas is about 4 $\times$10$^{{\rm 8}}$ M$_{\odot}$, which is
considerable, but only a small fraction of the gaseous content of the
Galaxy, and orders of magnitude less than the values suggested for the
Local Group models \citep{nica02}.  This amount of mass is similar to that found in the
less X-ray luminous early-type galaxies, and it is comparable to the amount of
mass suggested to be present due to Galactic Fountain models \citep{
rosen95,avill04}.  Also, this mass is much less than the external baryon mass
predicted by in some N-body/gasdynamical simulations \citep{somm06}, which
is expected to be comparable to the baryonic mass of the disk plus bulge stars
and the gas in the Galaxy (5 $\times$ 10$^{{\rm 10}}$ M$_{\odot}$).  
While we have assumed a Solar-like metallicity, we note
that the oxygen abundance cannot be significantly lower, as it would lead to a
higher electron column density that would exceed the values
determined from the dispersion measure toward pulsars.  At the pole, the
pulsar dispersion measure implies an electron column of 5.1 $\times$ 10$^{{\rm 19}}$
cm$^{-2}$ \citep{taylor93}, comparable to the column inferred from the
\ion{O}{7}.  Already, there is hardly room for the contribution from the 10$^{{\rm 4}}$ K gas
that is known to exist around the disk \citep{reyn04}, so any decrease in the
oxygen abundance or the \ion{O}{7} fraction can be excluded.  The constraint
provided by the pulsar dispersion measure argues for near-Solar
abundances and an \ion{O}{7} ionization near unity.

A weak limit on the size of hot halos around other galaxies is provided by
the failure to find enhanced absorption toward M31, where the sightline
of the nearest object passes 380 kpc from its center.  Also, there was no
substantial enhancement in the \ion{O}{7} (or \ion{O}{8}) absorption toward LMC
X-3, which suggests that there is not a significant 10$^{{\rm 6}}$ K halo around the
LMC.  This is consistent with the results of \citet{wang05}, who showed
that the redshift of the line center, as determined with {\it Chandra\/}, is
inconsistent with an LMC origin and consistent with a Galactic origin.  A
cooler ionized halo is known to exist around the LMC, which is
responsible for the enhanced dispersion measure toward pulsars \citep{manc06}.

This study does not rule out some component
of the \ion{O}{7} emission lying in a Local Group medium.  Certainly, the
Local Group is not devoid of gas, so it must have some \ion{O}{7} as well. 
However, as \citet{fang05b} discuss, if galaxy groups commonly had a
significant \ion{O}{7} column, it would be detected in sight lines toward
background AGNs at a rate in conflict with existing limits.

There is more work that can be carried out with {\it XMM\/} data, as new
observations are improving the S/N for some AGNs and additional AGNs
are being observed for significant amounts of time.  However, {\it XMM\/} and
{\it Chandra\/} will not be able to provide us with $\sim$ 10$^{{\rm 2}}$ sight lines with high
S/N line measurements, which would be needed to map out the structure
of the hot halo, much in the same way as has been accomplished for lower
temperature gas with {\it IUE\/}, {\it HST\/}, and {\it FUSE\/} spectra.  An improvement in
the X-ray capabilities of an order of magnitude would provide the
necessary data, and this will become possible with {\it Constellation-X\/}.

\acknowledgements
The authors would like to thank Andy Rasmussen, Wilt Sanders, Jimmy Irwin,
and Renato Dupke for their comments and insight.  We are indebted to those who
have developed and maintained the XMM archive, upon which this work was based.
Also, we would like to acknowledge support from NASA for these activities,
through the Long Term Space Astrophysics grant NAG5-10765, and also NAG5-13137.

\clearpage

\begin{deluxetable}{rlrrcrr}
\tabletypesize{\scriptsize}
\tablecaption{Absorption Line Targets: Basic Data}
\tablewidth{0pt}
\tablehead{
\colhead{No.} & \colhead{Name} & \colhead{\it{l}} & \colhead{\it{b\/}}
 & \colhead{z} & \colhead{N$_{H}$} & \colhead{t$_{exp}$\tablenotemark{a}} \\ 
\colhead{} & \colhead{} & \colhead{(deg)} & \colhead{(deg)} & 
 \colhead{} & \colhead{(10$^{20}$cm$^{-2}$}) & \colhead{(ksec)} 
}
\startdata
1 & Mkn 421 & 179.83 & 65.03 & 0.0300 & 1.38 & 509 \\ 
2 & PKS 2155-304 & 17.73 & -52.24 & 0.116 & 1.71 & 723 \\ 
3 & 3C 273 & 289.95 & 64.36 & 0.1583 & 1.79 & 451 \\ 
4 & MCG-6-30-15 & 313.29 & 27.68 & 0.0077 & 4.06 & 460 \\ 
5 & LMC X-3 & 273.57 & -32.08 & 0.000927 & 4.74 & 126 \\ 
6 & 1H1426+428 & 77.49 & 64.90 & 0.129 & 1.38 & 388 \\ 
7 & Akn 564 & 92.14 & -25.34 & 0.0247 & 6.27 & 130 \\ 
8 & NGC 4051 & 148.88 & 70.09 & 0.0023 & 1.32 & 174 \\ 
9 & NGC 3783 & 287.46 & 22.95 & 0.0097 & 8.26 & 316 \\ 
10 & NGC 5548 & 31.96 & 70.50 & 0.0172 & 1.69 & 160 \\ 
11 & Akn120 & 201.69 & -21.13 & 0.0327 & 12.2 & 112 \\ 
12 & PKS 0558-504 & 257.96 & -28.57 & 0.137 & 4.39 & 168 \\ 
13 & Mkn766 & 190.68 & 82.27 & 0.0129 & 1.71 & 189 \\ 
14 & NGC 4593 & 297.48 & 57.40 & 0.009 & 2.31 & 115 \\ 
15 & 3C 390.3 & 111.44 & 27.07 & 0.0561 & 4.24 & 123 \\ 
16 & NGC 7469 & 83.10 & -45.47 & 0.0163 & 4.87 & 189 \\ 
17 & Mkn 509 & 35.97 & -29.86 & 0.0344 & 4.11 & 75 \\ 
18 & 3C 120 & 190.37 & -27.40 & 0.033 & 11.1 & 129 \\ 
19 & NGC 3516 & 133.24 & 42.40 & 0.0088 & 3.05 & 260 \\ 
20 & Ton 1388 & 223.36 & 68.21 & 0.1765 & 1.28 & 110 \\ 
21 & 1H 0414+009 & 191.82 & -33.16 & 0.287 & 10.3 & 79 \\ 
22 & Mr2251-178 & 46.20 & -61.33 & 0.064 & 2.70 & 65 \\ 
23 & IC 4329a & 317.5 & 30.92 & 0.0161 & 4.42 & 136 \\ 
24 & Fairall 9 & 295.07 & -57.83 & 0.047 & 3.19 & 33 \\ 
25 & MS0737.9+7441 & 140.27 & 29.57 & 0.315 & 3.54 & 86 \\ 
26 & 3C 59 & 142.04 & -30.54 & 0.1096 & 5.84 & 82 \\ 
\enddata
\tablenotetext{a}{Cleaned exposure time for the {\it XMM RGS}.}
\end{deluxetable}
\clearpage

\begin{deluxetable}{rlrrrrrrrrrrrr}
\tabletypesize{\scriptsize}
\rotate
\tablecaption{OVII Absorption Line Measurements}
\tablewidth{0pt}
\tablehead{
\colhead{No.} & \colhead{Name} & \colhead{AngleGC} & \colhead{AngleM31}
 & \colhead{R12} & \colhead{R45} & \colhead{R67} & \colhead{F$_{\lambda}$}
 & \colhead{S/N} & \colhead{${\lambda}_c$} & \colhead{2$\sigma$} & \colhead{EW} 
 & \colhead{err} & \colhead{S/N} \\ 
\colhead{} & \colhead{} & \colhead{(deg)} & \colhead{(deg)} & \colhead{} & \colhead{}
 & \colhead{} &  \colhead{}
 & \colhead{} & \colhead{(\AA )} & \colhead{(m\AA )}
& \colhead{(m\AA )} & \colhead{(m\AA )} & \colhead{} 
}
\startdata
1 & Mkn 421 & 114.97 & 97.42 & 1087 & 105 & 116 & 1.24E-02 & 48.9 & 21.602 & 0.069 & 11.8 & 0.8 & 14.1 \\ 
2 & PKS 2155-304 & 54.32 & 80.89 & 964 & 145 & 135 & 2.88E-03 & 26.5 & 21.605 & 0.081 & 13.7 & 1.9 & 7.4 \\ 
3 & 3C 273 & 81.51 & 136.57 & 1374 & 189 & 132 & 1.43E-03 & 14.7 & 21.608 & 0.083 & 24.6 & 3.3 & 7.4 \\ 
4 & MCG-6-30-15 & 52.61 & 167.41 & 826 & 304 & 189 & 9.42E-04 & 11.2 & 21.617 & 0.081 & 32.6 & 6.8 & 4.8 \\ 
5 & LMC X-3 & 86.97 & 120.20 & 533 & 160 & 156 & 9.00E-03 & 8.9 & 21.592 & 0.063 & 21.0 & 5.0 & 4.2 \\ 
6 & 1H1426+428 & 84.73 & 92.73 & 1084 & 124 & 114 & 7.74E-04 & 8.0 & 21.603 & 0.08 & 11.6 & 4.1 & 2.8 \\ 
7 & Akn 564 & 91.93 & 26.84 & 429 & 115 & 115 & 1.85E-03 & 7.4 & 21.603 & 0.08 & 12.3 & 4.6 & 2.7 \\ 
8 & NGC 4051 & 106.95 & 93.74 & 1118 & 115 & 118 & 7.20E-04 & 7.4 & 21.603 & 0.05 & 24.6 & 3.1 & 7.9 \\ 
9 & NGC 3783 & 73.97 & 167.24 & 510 & 143 & 118 & 3.52E-04 & 5.9 & 21.603 & 0.08 & 24.1 & 7.6 & 3.2 \\ 
10 & NGC 5548 & 73.54 & 110.02 & 1052 & 152 & 139 & 7.79E-04 & 5.7 & 21.603 & 0.08 & 7.0 & 6.8 & 1.0 \\ 
11 & Akn120 & 150.08 & 74.02 & 444 & 141 & 155 & 7.03E-04 & 5.6 & 21.603 & 0.08 & -6.0 & 5.5 & -1.1 \\ 
12 & PKS 0558-504 & 100.56 & 114.80 & 500 & 122 & 120 & 6.89E-04 & 5.1 & 21.613 & 0.05 & 21.7 & 7.8 & 2.8 \\ 
13 & Mkn766 & 97.60 & 108.70 & 887 & 112 & 112 & 8.33E-04 & 5.1 & 21.603 & 0.08 & 0.2 & 6.8 & 0.0 \\ 
14 & NGC 4593 & 75.61 & 144.07 & 861 & 201 & 153 & 9.46E-04 & 5.0 & 21.603 & 0.08 & 23.4 & 8.5 & 2.8 \\ 
15 & 3C 390.3 & 108.99 & 49.55 & 512 & 138 & 125 & 4.24E-04 & 4.3 & 21.603 & 0.08 & 27.4 & 7.3 & 3.8 \\ 
16 & NGC 7469 & 85.17 & 39.15 & 406 & 100 & 117 & 6.96E-04 & 4.2 & 21.603 & 0.08 & 1.6 & 8.9 & 0.2 \\ 
17 & Mkn 509 & 45.42 & 75.49 & 743 & 220 & 147 & 6.23E-04 & 4.0 & 21.603 & 0.08 & 25.9 & 7.3 & 3.6 \\ 
18 & 3C 120 & 150.85 & 62.46 & 466 & 111 & 117 & 2.81E-04 & 3.5 & 21.603 & 0.08 & 13.8 & 9.2 & 1.5 \\ 
19 & NGC 3516 & 120.39 & 64.94 & 831 & 113 & 118 & 9.04E-05 & 2.5 & 21.603 & 0.08 & 22.0 & 13.4 & 1.6 \\ 
20 & Ton 1388 & 105.66 & 114.48 & 1004 & 121 & 128 & 1.41E-04 & 2.3 & 21.603 & 0.08 & 34.5 & 15.7 & 2.2 \\ 
21 & 1H 0414+009 & 145.03 & 62.68 & 612 & 141 & 144 & 1.66E-04 & 2.2 & 21.603 & 0.08 & -3.1 & 14.8 & -0.2 \\ 
22 & Mr2251-178 & 70.60 & 64.00 & 646 & 137 & 126 & 1.37E-04 & 1.8 & 21.603 & 0.08 & 39.8 & 19.6 & 2.0 \\ 
23 & IC 4329a & 50.77 & 162.66 & 669 & 248 & 155 & 1.85E-04 & 1.7 & 21.603 & 0.08 & 33.8 & 19.3 & 1.8 \\ 
24 & Fairall 9 & 76.96 & 100.44 & 893 & 122 & 124 & 2.17E-04 & 1.7 & 21.603 & 0.08 & 31.1 & 16.5 & 1.9 \\ 
25 & MS0737.9+7441 & 131.98 & 54.35 & 548 & 112 & 116 & 1.30E-04 & 1.6 & 21.603 & 0.08 & -13.9 & 20.7 & -0.7 \\ 
26 & 3C 59 & 132.77 & 20.73 & 493 & 128 & 115 & 9.22E-05 & 1.5 & 21.603 & 0.08 & 60.9 & 19.2 & 3.2 \\ 
\enddata
\tablecomments{AngleGC is the angle from the Galactic Center; AngleM31 is the angle from M31;
 R12, R45, and R67 are the {\it ROSAT} soft X-ray background values in units of 
10$^{-6}$ cnt s$^{-1}$ arcmin$^{-2}$; F$_{\lambda}$ is the continuum flux 
at 21.60 \AA\ in units of  cnt s$^{-1}$ {\AA }$^{-1}$ cm$^{-2}$, followed by its S/N;
 ${\lambda}_c$ the the line center of the OVII absorption; $\sigma$ is 
the Gaussian parameter describing the OVII line width; 
and EW is the equivalent width of the line.
}
\end{deluxetable}

\clearpage

\begin{figure}
\plotone{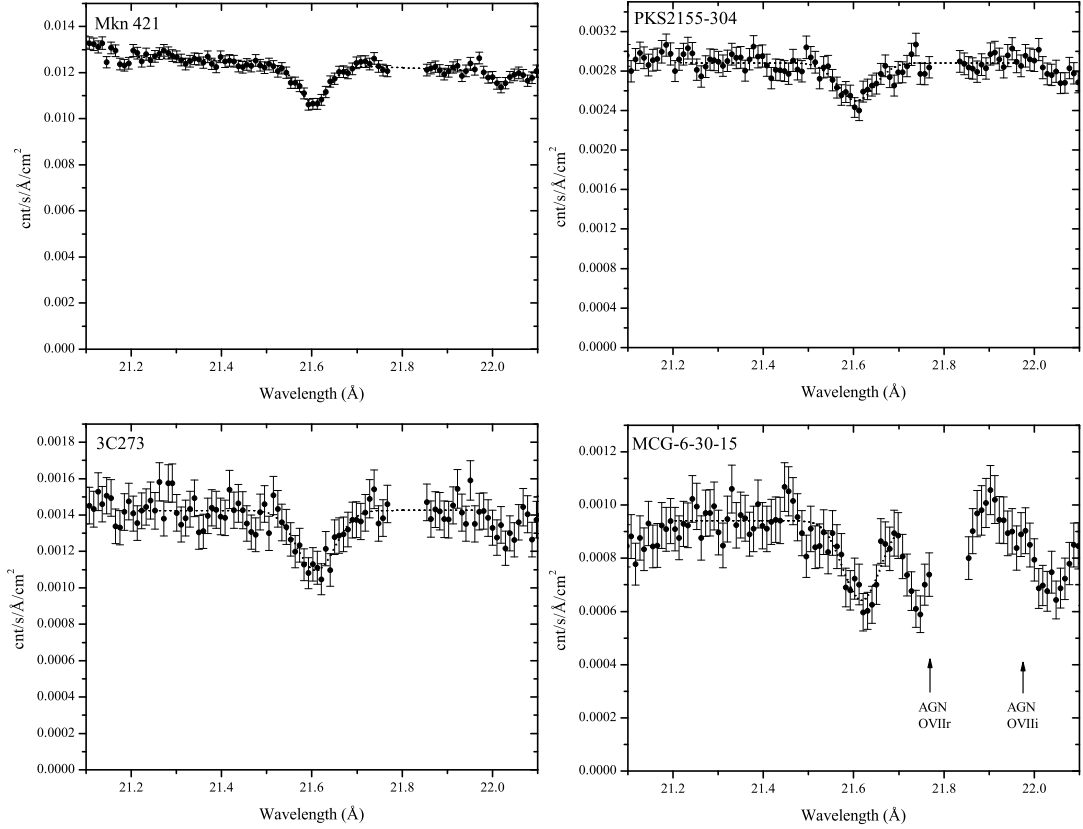}
\caption{The {\it XMM} flux, corrected for instrumental features, at the 
location of the  \ion{O}{7} line at 21.60 \AA .  There is an instrumental
feature near 21.82 \AA , which is not used in the fit; the fit is shown by the dotted
line.  A feature near 22.0 \AA\ is from a Galactic \ion{O}{6} blend.  The
objects are shown by decreasing S/N of the continuum.  For MCG-6-30-15, the
location of the resonance and intercombination lines at the redshift of
the object is shown.}
\end{figure}

\clearpage

\begin{figure}
\plotone{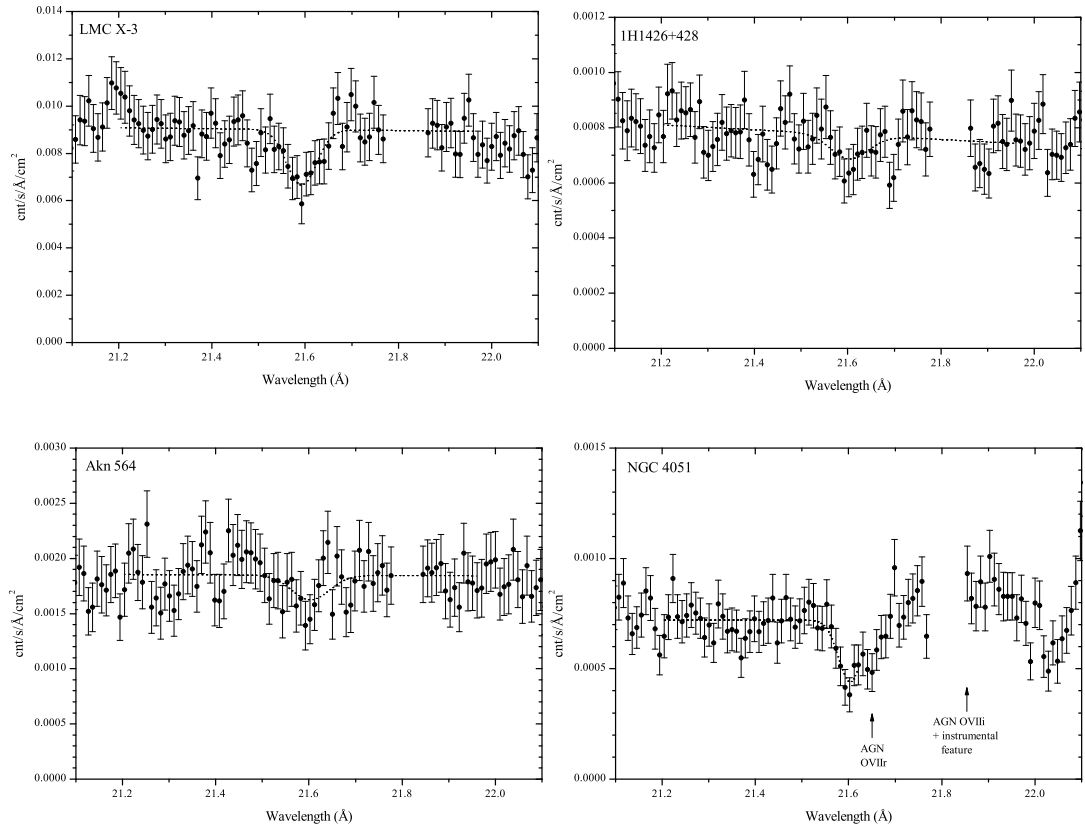}
\caption{As above, where the objects are ordered by continuum S/N.  The
location of the AGN \ion{O}{7} lines in NGC 4051, if present are shown.  For this
source, the continuum is defined by the data blueward of the 21.60 \AA\ line.}
\end{figure}

\clearpage

\begin{figure}
\plotone{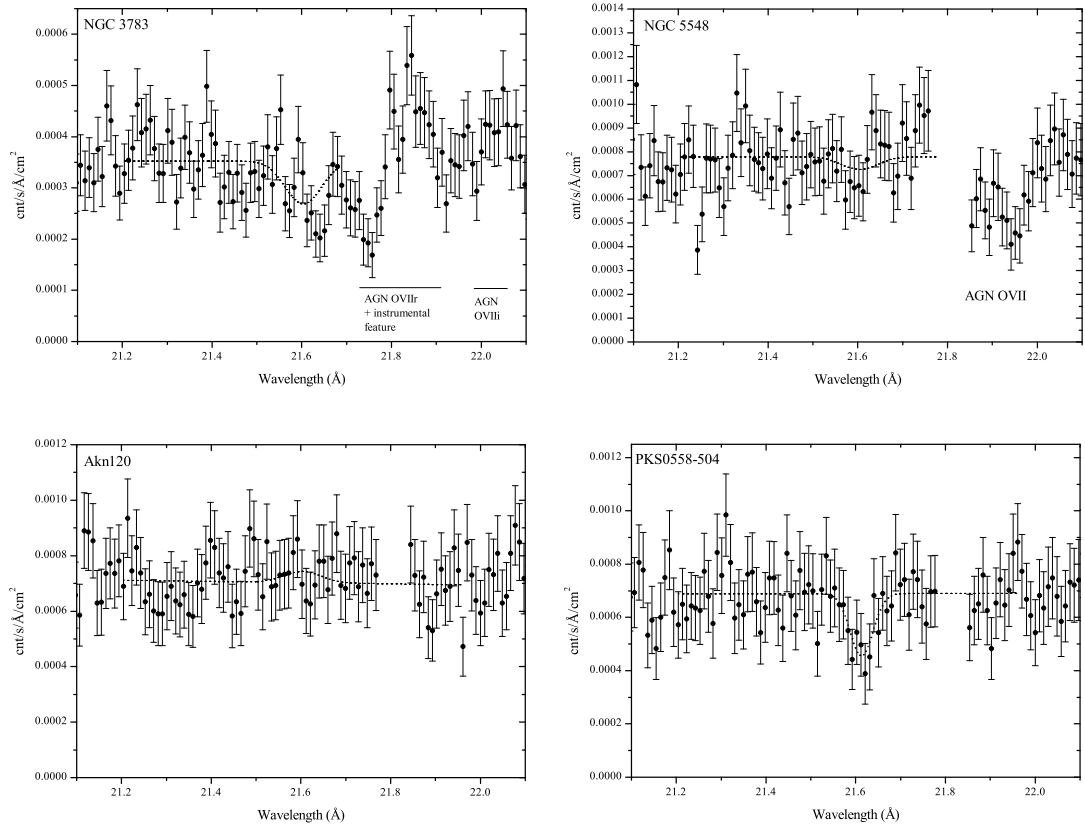}
\caption{As above, for targets 9-12.  NGC 3783 has a possible P Cygni profile
in its redshifted resonance \ion{O}{7} line.}
\end{figure}

\clearpage

\begin{figure}
\plotone{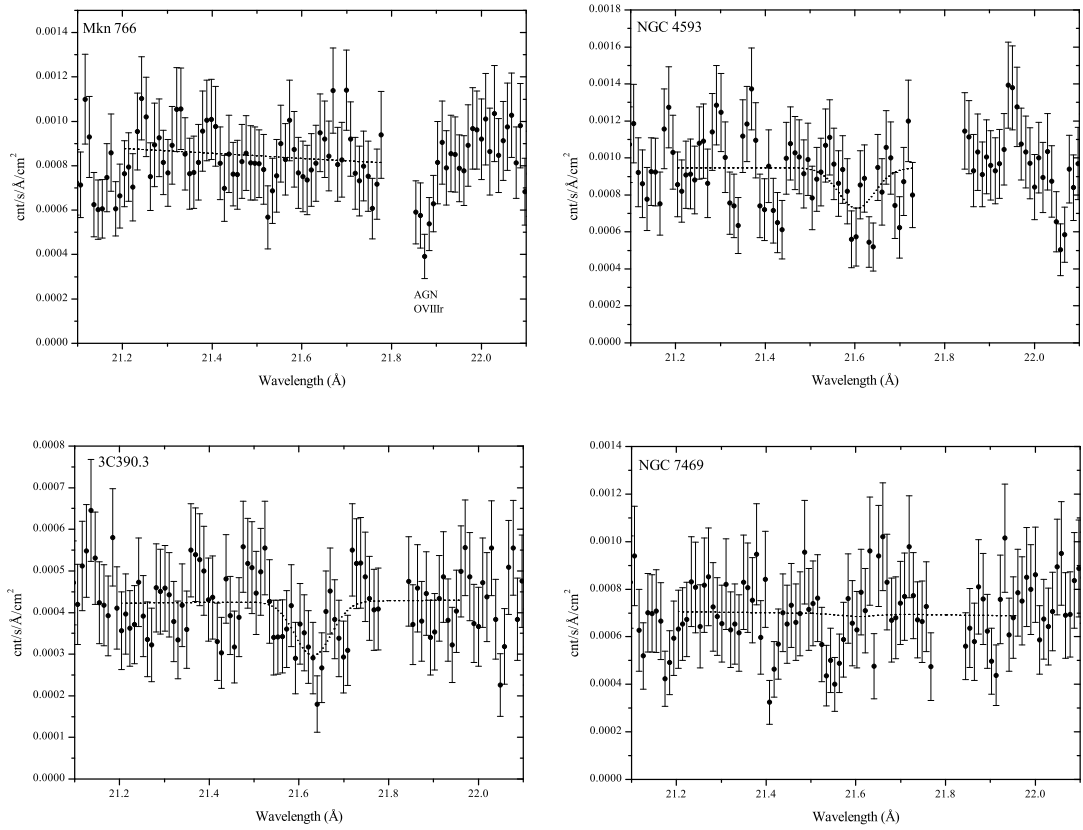}
\caption{As above, for targets 13-16. }
\end{figure}

\clearpage

\begin{figure}
\plotone{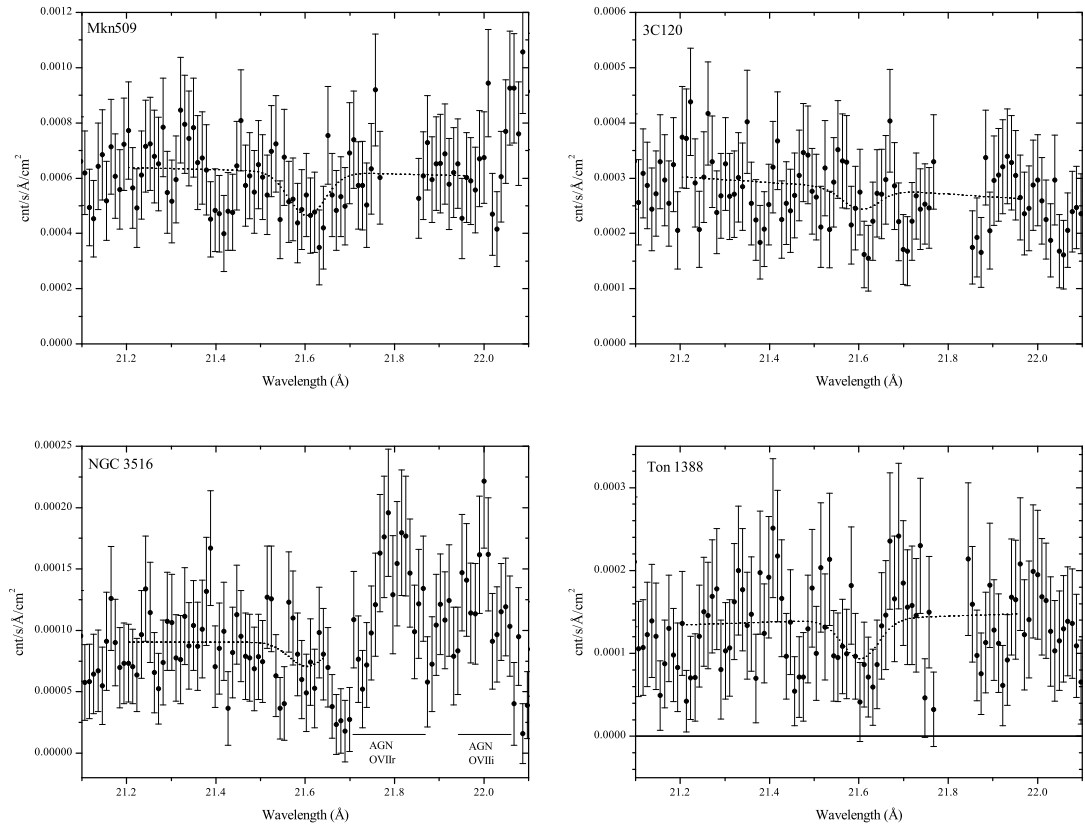}
\caption{As above, for targets 17-20.  For most of the analysis, we do not
use targets 19 and higher and there is a decrease in S/N such that the 
uncertainty in the equivalent widths become too large.  In NGC 3516, both
the redshifted resonance and intercombination lines are present in emission.}
\end{figure}

\clearpage

\begin{figure}
\plotone{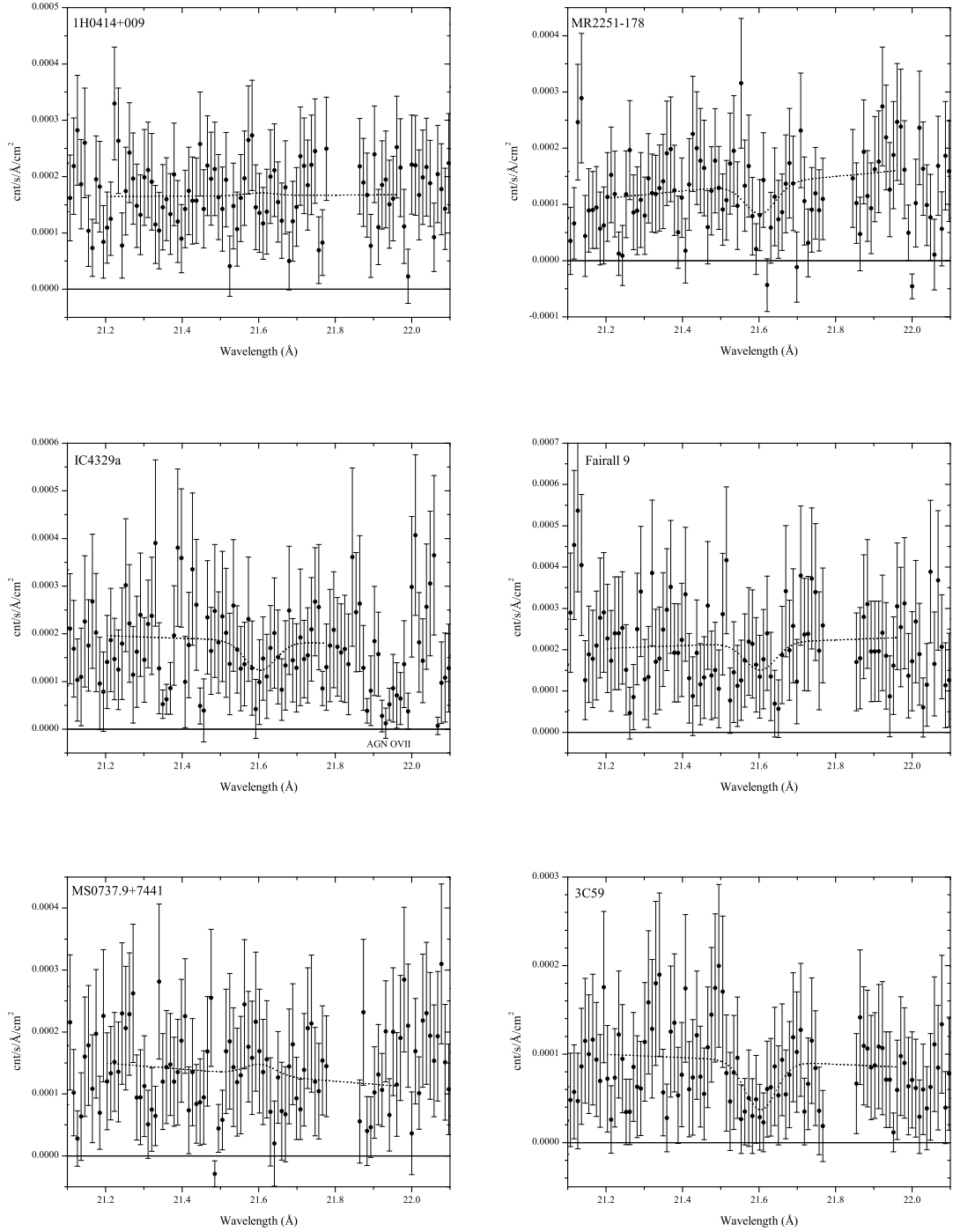}
\caption{For these final six objects, the S/N is rather poor.  The 
faintest source, 3C59, which shows possible absorption, lies within
130 kpc of M33. }
\end{figure}

\clearpage

\begin{figure}
\plotone{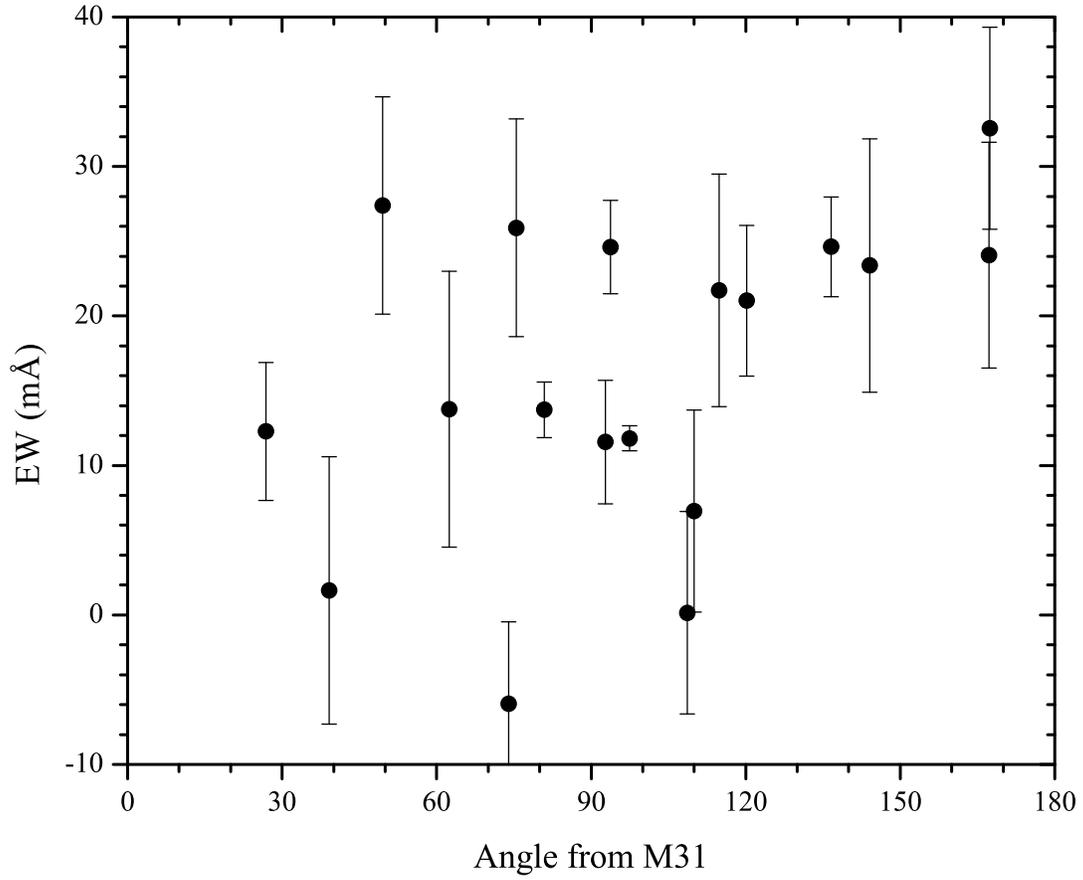}			
\caption{The distribution of \ion{O}{7} equivalent widths as a function
of angle from M31.  In the Local Group model, we would expect a dereasing
equivalent width with increasing angle, a prediction not verified. }
\end{figure}

\clearpage

\begin{figure}
\plotone{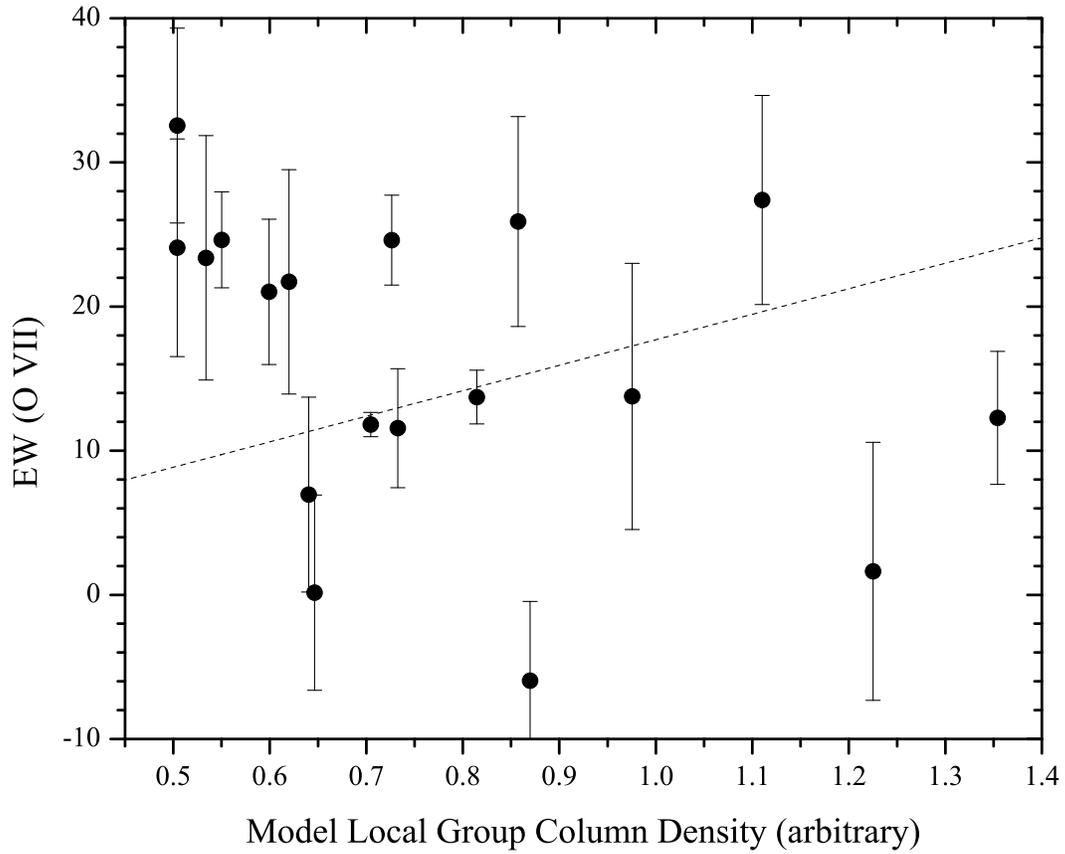}
\caption{The observed \ion{O}{7} equivalent widths as a function
of the column density of an ellipse of diffuse gas filling the Local Group
and oriented along the Milky Way - M31 axis.  The data are in conflict with
the predicted distribution, shown as a dashed line. }
\end{figure}

\clearpage

\begin{figure}
\plotone{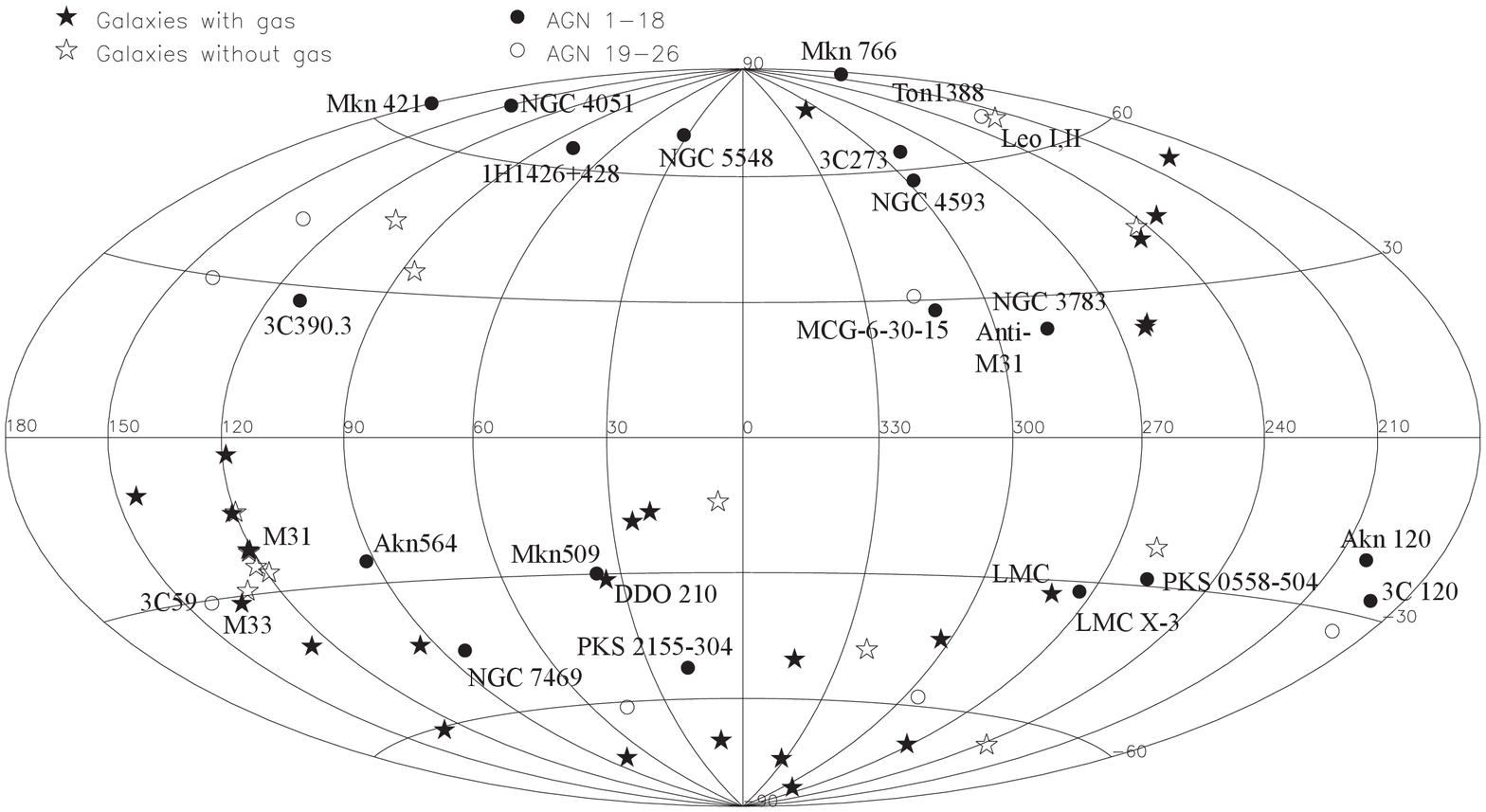}
\caption{The distribution in the sky of the X-ray absorption sources 
(black circles are objects 1-18 and are labeled; open circles are objects 19-26)
is shown along with the Local Group galaxies (stars; black indicates HI or
H$_2$ gas was detected).  Some of the prominent galaxies are labeled as are
any moderatly close galaxy-AGN pairs; the anti-M31 direction is labeled.}
\end{figure}

\clearpage

\begin{figure}
\plotone{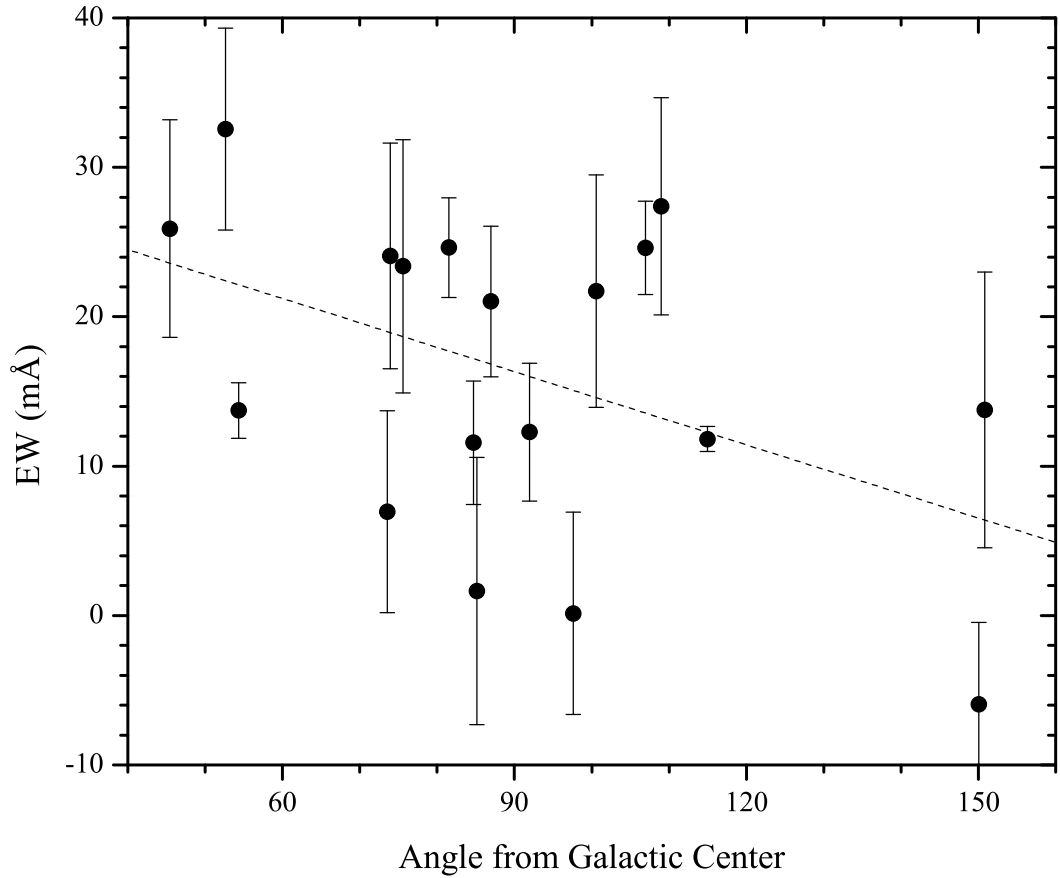}			
\caption{The \ion{O}{7} equivalent widths as a function of the angle from
the Galactic Center, which shows a trend that is consistent with the
Galactic Halo model.  The dotted line is the best-fit, which is significant 
at the 95\% level. }
\end{figure}

\clearpage

\begin{figure}
\plotone{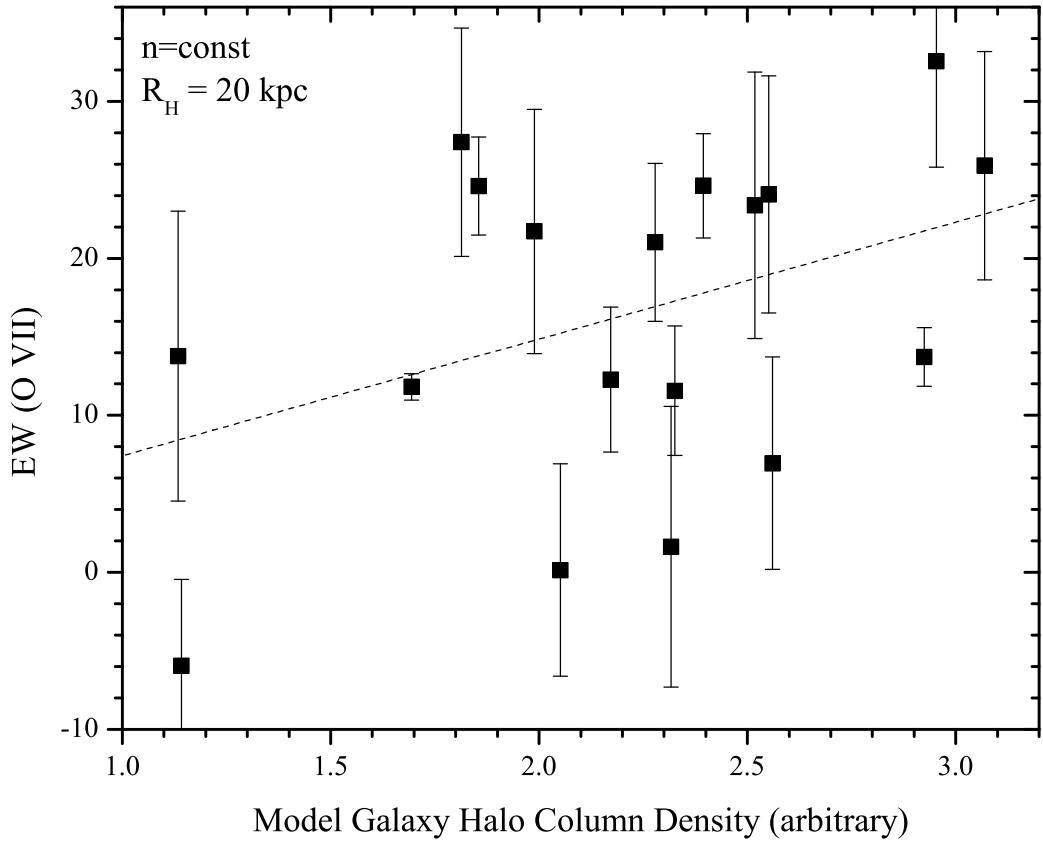}
\caption{The \ion{O}{7} equivalent widths as a function of the column density
for a Galactic halo of uniform density and radius of 20 kpc as measured from
the Galactic Center.  The model line passes through the origin and the fit 
is significant at the 95\% level. }
\end{figure}

\clearpage

\begin{figure}
\plotone{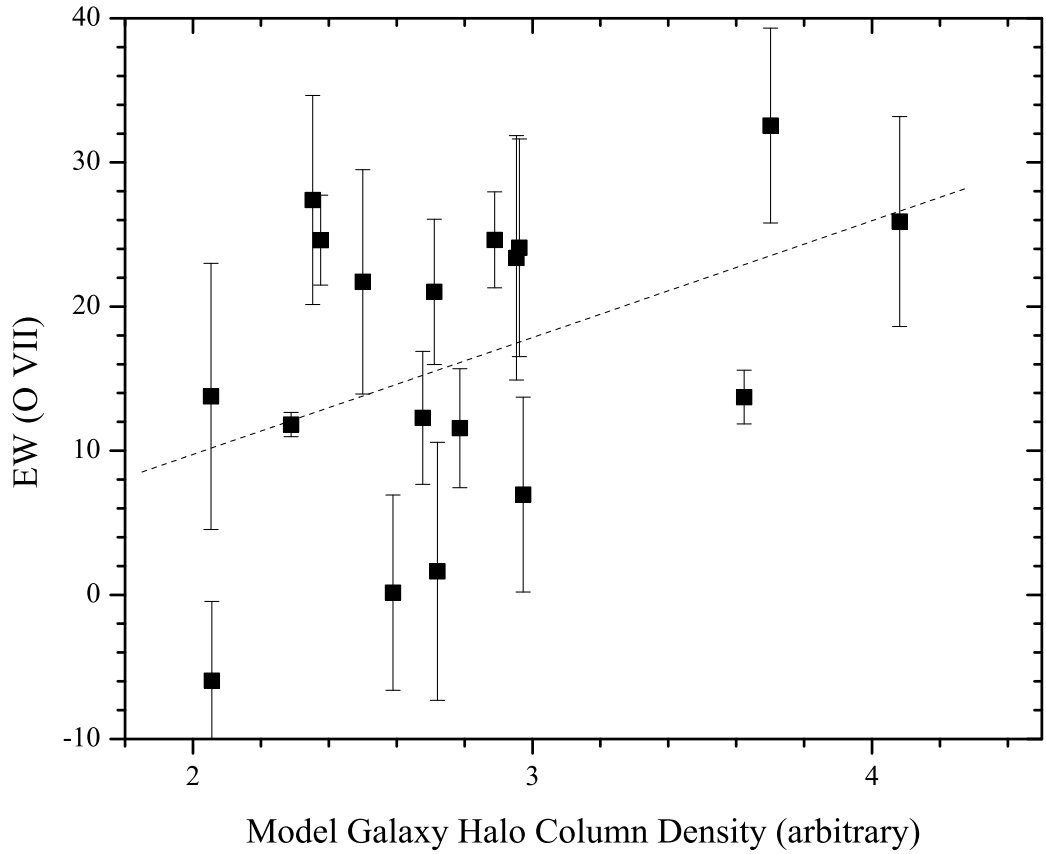}
\caption{The \ion{O}{7} equivalent widths as a function of the column density
for a Galactic halo where the density n $\propto$ r$^{-3/2}$.  The fit shown, 
which is not required to go through the origin is significant at the 95\% level. }
\end{figure}

\clearpage

\begin{figure}
\plotone{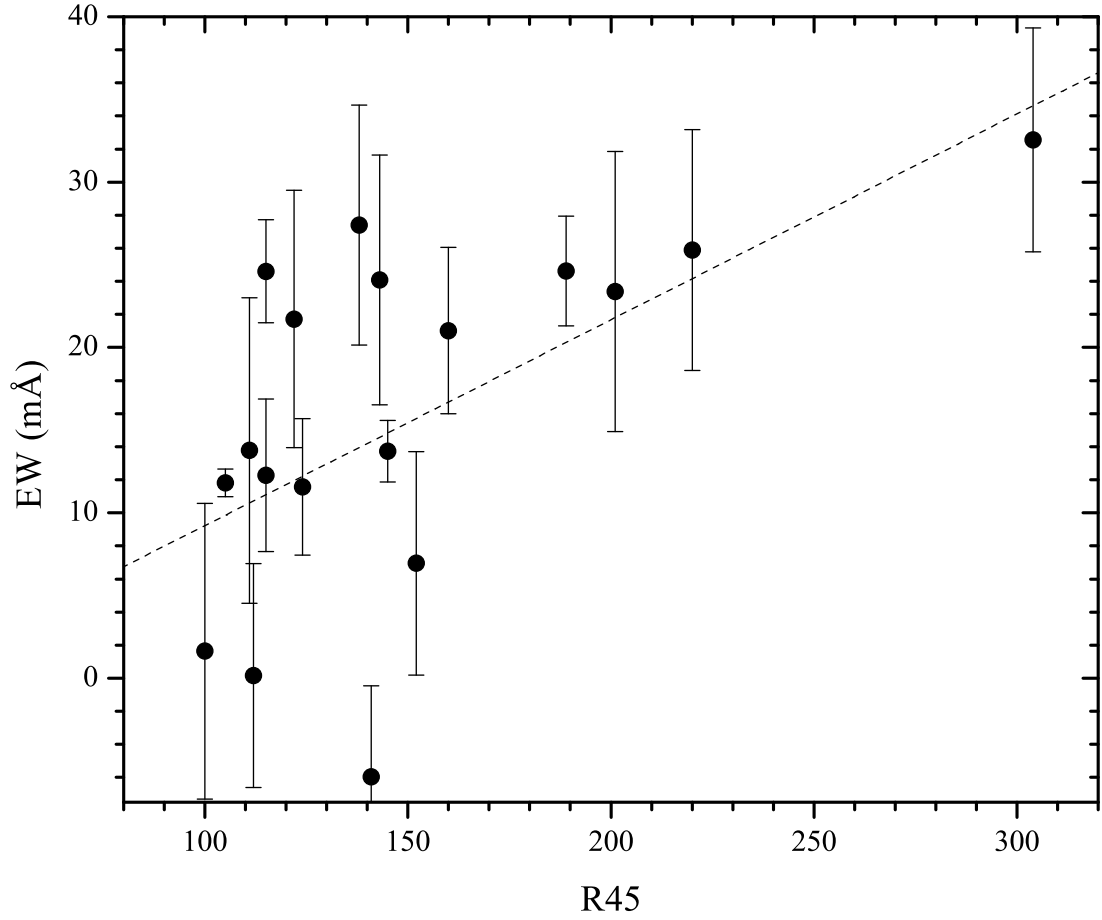}		
\caption{The \ion{O}{7} equivalent widths as a function of the R45 intensity in
an annulus around each target; the units for R45 are 10$^{-6}$ counts s$^{-1}$ arcmin$^{-2}$.
The fit (dashed line) is significant at the 98-99\% confidence level.}
\end{figure}

\clearpage

\begin{figure}
\plotone{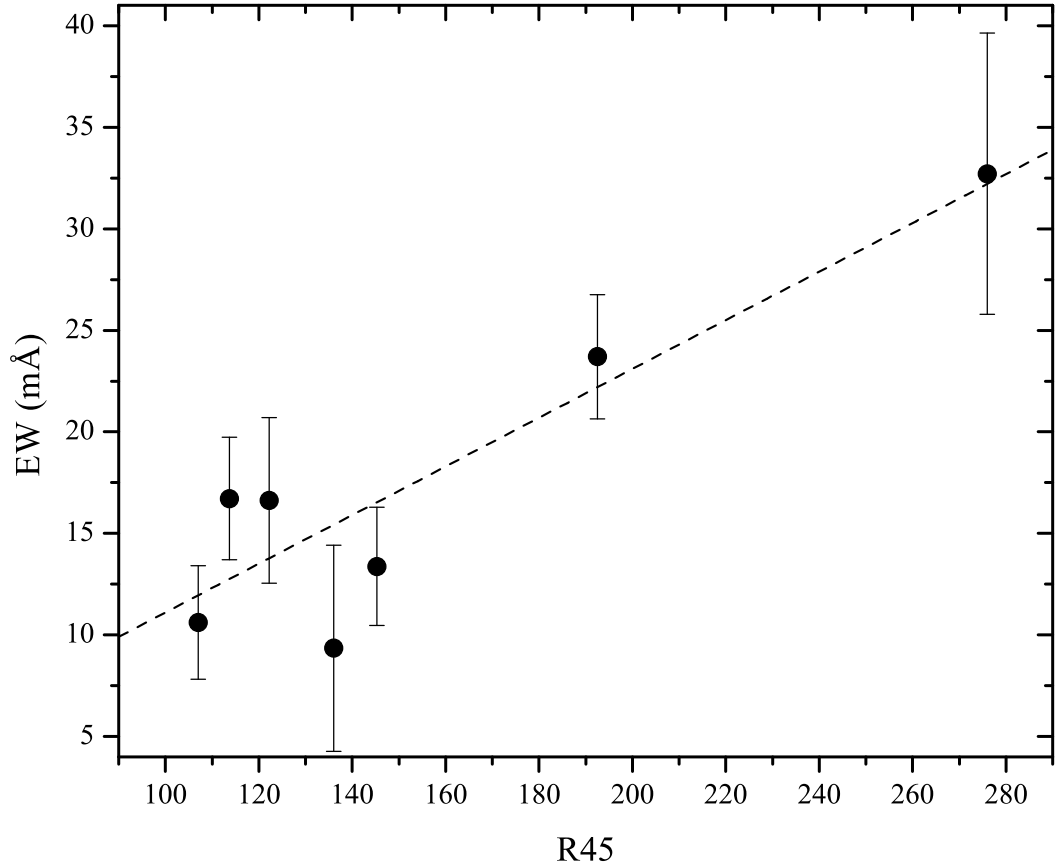}
\caption{As above, except that all 26 objects are used to make binned, 
error-weighted averages.  We have incorporated an additional rms to the 
equivalent widths of 3 m\AA\ so that the $\chi$$^{{\rm 2}}$ of the error-weighted fit 
is in the acceptable range. The correlation is significant at the 
99\% confidence level.}
\end{figure}



\end{document}